\documentstyle{mn}
\input psfig.sty
\input{epsf}

\def \xoff {\ifmmode x_{\rm off} \else $x_{\rm off}$ \fi}
\def \rhorms {\ifmmode \rho_{\rm rms} \else $\rho_{\rm rms}$ \fi}

\def \apj  {ApJ}
\def \apjl  {ApJL}
\def \aj  {AJ}
\def\mnras {MNRAS}
\def \etal {et~al.~}
\def \chisq  {\ifmmode  \chi^2   \else  $\chi^2$  \fi}  
\def \chisqr {\ifmmode \chi^2_{\rm r} \else $\chi^2_{\rm r}$ \fi}
\def \spose#1{\hbox  to 0pt{#1\hss}}  
\def \lta{\mathrel{\spose{\lower 3pt\hbox{$\sim$}}\raise  2.0pt\hbox{$<$}}}
\def \gta{\mathrel{\spose{\lower  3pt\hbox{$\sim$}}\raise 2.0pt\hbox{$>$}}}
 
\def \ha  {\ifmmode H\alpha \else H$\alpha $ \fi}

\def \kms {\ifmmode  \,\rm km\,s^{-1} \else $\,\rm km\,s^{-1}  $ \fi }
\def \kpc {\ifmmode  {\rm kpc}  \else ${\rm  kpc}$ \fi  }  
\def \Msun {\ifmmode M_{\odot} \else $M_{\odot}$ \fi} 
\def \hMsun {\ifmmode h^{-1}\,\rm M_{\odot} \else $h^{-1}\,\rm M_{\odot}$ \fi}
\def \hhMsun {\ifmmode h^{-2}\,\rm M_{\odot}\else $h^{-2}\,\rm M_{\odot}$ \fi}
\def \Lsun {\ifmmode L_{\odot} \else $L_{\odot}$ \fi} 
\def \hhLsun {\ifmmode h^{-2}\,\rm L_{\odot} \else $h^{-2}\,\rm L_{\odot}$ \fi}

\def \LCDM {\ifmmode \Lambda{\rm CDM} \else $\Lambda{\rm CDM}$ \fi}
\def \sig8 {\ifmmode \sigma_8 \else $\sigma_8$ \fi} 
\def \OmegaM {\ifmmode \Omega_{\rm M} \else $\Omega_{\rm M}$ \fi} 
\def \OmegaL {\ifmmode \Omega_{\rm \Lambda} \else $\Omega_{\rm \Lambda}$\fi} 
\def \Deltavir {\ifmmode \Delta_{\rm vir} \else $\Delta_{\rm vir}$ \fi}

\def \rs {\ifmmode r_{\rm s} \else $r_{\rm s}$ \fi} 
\def \rrm2 {\ifmmode r_{-2} \else $r_{-2}$ \fi} 
\def \ccm2 {\ifmmode c_{-2} \else$c_{-2}$ \fi} 
\def \cvir {\ifmmode c_{\rm vir} \else $c_{\rm vir}$ \fi} 
\def \cbar {\ifmmode \overline{c} \else $\overline{c}$ \fi}

\def \R200 {\ifmmode R_{200} \else $R_{200}$ \fi} 
\def \Rvir {\ifmmode R_{\rm vir} \else $R_{\rm vir}$ \fi}

\def \v200 {\ifmmode V_{200} \else $V_{200}$ \fi} 
\def \Vvir {\ifmmode V_{\rm  vir} \else  $V_{\rm vir}$  \fi} 
\def  \Vhalo  {\ifmmode V_{\rm halo} \else $V_{\rm halo}$ \fi}

\def \M200 {\ifmmode M_{200} \else $M_{200}$ \fi} 
\def \Mvir {\ifmmode M_{\rm  vir} \else $M_{\rm  vir}$ \fi}  
\def \Mshell  {\ifmmode M_{\rm shell} \else $M_{\rm shell}$ \fi}

\def \Nvir {\ifmmode N_{\rm  vir} \else $N_{\rm  vir}$ \fi}  

\def \Jvir {\ifmmode J_{\rm vir} \else $J_{\rm vir}$ \fi} 
\def \Jshell {\ifmmode J_{\rm shell} \else $J_{\rm shell}$ \fi}

\def \Evir {\ifmmode E_{\rm vir} \else $E_{\rm vir}$ \fi} 

\def \lam {\ifmmode \lambda  \else $\lambda$ \fi} 
\def \lamp {\ifmmode \lambda^{\prime} \else $\lambda^{\prime}$  \fi} 
\def \lampc {\ifmmode \lambda^{\prime}_{\rm c} \else
  $\lambda^{\prime}_{\rm c}$  \fi} 
\def \lambar {\ifmmode \bar{\lambda}  \else  $\bar{\lambda}$  \fi}  
\def  \lampbar  {\ifmmode \bar{\lambda^{\prime}} \else
  $\bar{\lambda^{\prime}}$\fi} 
\def \siglam {\ifmmode \sigma_{\lambda} \else $\sigma_{\lambda}$ \fi} 
\def \siglamp {\ifmmode                \sigma_{\lambda^{\prime}} \else
$\sigma_{\lambda^{\prime}}$\fi}

\def \Rd {\ifmmode R_{\rm d} \else $R_{\rm d}$ \fi} 
\def \Rs {\ifmmode R_{\rm s} \else $R_{\rm s}$ \fi}  
\def \Rd {\ifmmode R_{\rm d} \else $R_{\rm d}$ \fi}  
\def \Rcool  {\ifmmode R_{\rm  cool}  \else $R_{\rm cool}$ \fi} 
\def \RIII {\ifmmode  3.2\Rs \else $3.2\Rs$ \fi} 
\def \RII {\ifmmode 2.2\Rs \else $2.2\Rs$  \fi} 
\def \Reff {\ifmmode R_{\rm eff} \else $R_{\rm  eff}$ \fi} 
\def  \rb {\ifmmode r_{\rm b}  \else $r_{\rm b}$ \fi}

\def  \Sigmacrit   {\ifmmode  \Sigma_{\rm  crit}   
\else  $\Sigma_{\rm crit}$\fi} 
\def \Sig0 {\ifmmode \Sigma_{0} \else $\Sigma_{0}$ \fi}

\def \muI {\ifmmode \mu_{0,I} \else $\mu_{0,I}$ \fi}

\def \mgal {\ifmmode m_{\rm gal} \else $m_{\rm gal}$ \fi} 
\def \md {\ifmmode m_{\rm d} \else $m_{\rm d}$ \fi} 
\def \ms {\ifmmode m_{\rm   s}   \else   $m_{\rm   s}$   \fi}   
\def   \mdbar   {\ifmmode {\overline{m}}_{\rm d} \else
  ${\overline{m}}_{\rm d}$ \fi} 
\def \msbar {\ifmmode  \bar{m}_{\rm  s}  \else  $\bar{m}_{\rm s}$
  \fi}  
\def  \Md {\ifmmode M_{\rm d}  \else $M_{\rm d}$ \fi} 
\def  \Ms {\ifmmode M_{\rm s} \else $M_{\rm  s}$ \fi} 
\def \Mb {\ifmmode  M_{\rm b} \else $M_{\rm b}$ \fi} 
\def \Mstar {\ifmmode  M_{\rm star} \else $M_{\rm star}$ \fi}
\def \Mdisc {\ifmmode M_{\rm disc} \else $M_{\rm disc}$ \fi}

\def \Jd {\ifmmode J_{\rm d} \else $J_{\rm d}$ \fi} 
\def \Jb {\ifmmode J_{\rm b} \else $J_{\rm b}$ \fi}  
\def \fb {\ifmmode  f_{\rm b} \else $f_{\rm b}$ \fi}

\def  \jd  {\ifmmode j_{\rm  d}  \else  $j_{\rm  d}$ \fi}  
\def  \jdmd {\ifmmode \frac{j_{\rm  d}}{m_{\rm d}} \else
  $\frac{j_{\rm d}}{m_{\rm d}}$ \fi} 
\def \fj {\ifmmode f_{\rm j} \else $f_{\rm j}$ \fi} 
\def \ft {\ifmmode f_{\rm t}  \else $f_{\rm t}$ \fi} 
\def  \fM {\ifmmode f_{\rm M} \else $f_{\rm M}$ \fi}

\def  \Vd {\ifmmode  V_{\rm  d}  \else $V_{\rm  d}$  \fi} 
\def  \Vcool {\ifmmode V_{\rm cool} \else $V_{\rm cool}$ \fi} 
\def \Vcirc {\ifmmode V_{\rm circ}  \else $V_{\rm circ}$  \fi} 
\def \VIII  {\ifmmode V_{3.2} \else $V_{3.2}$ \fi} 
\def  \VII {\ifmmode V_{2.2} \else $V_{2.2}$ \fi}
\def \Vobs {\ifmmode V_{\rm obs}  \else $V_{\rm obs}$ \fi} 
\def \Vdisc {\ifmmode V_{\rm disc} \else  $V_{\rm disc}$ \fi} 
\def \Vmax {\ifmmode V_{\rm  max} \else  $V_{\rm max}$  \fi} 
\def  \Vmaxobs{\ifmmode V_{\rm max}^{\rm obs}\else  $V_{\rm max}^{\rm
    obs}$\fi}  
\def \Vtot {\ifmmode V_{\rm tot} \else $V_{\rm tot}$  \fi} 
\def \Vrot {\ifmmode V_{\rm rot} \else  $V_{\rm rot}$  \fi} 
\def  \Vflat {\ifmmode  V_{\rm  flat} \else $V_{\rm flat}$ \fi}

\def \Ups {\ifmmode \Upsilon  \else $\Upsilon$ \fi} 
\def \YB {\ifmmode \Upsilon_B \else $\Upsilon_B$ \fi} 
\def \YI {\ifmmode  \Upsilon_I  \else $\Upsilon_I$ \fi} 
\def \DeltaIMF {\ifmmode \Delta_{\rm IMF} \else $\Delta_{\rm IMF}$ \fi}

\def\LCDM{$\Lambda$CDM }

\def\c200{$c_{200}$}

\title[Universal Substructure Distributions in $\Lambda$CDM halos]
      {Universal Substructure Distributions in $\Lambda$CDM halos: Can we find a Fossil Group?}
\author[E. D'Onghia et al.]
{Elena D'Onghia$^{1}$ \thanks{Marie Curie Fellow; elena@physik.unizh.ch},
Andrea V. Macci\`o$^{1,2}$, George Lake$^1$, Joachim Stadel$^1$, Ben Moore$^{1}$ \\
$^1$Institute for Theoretical Physics, University of Z\"urich,
Winterthurerstrasse 190 ,CH-8057 Z\"urich, Switzerland \\
$^2$Max-Planck-Institute for Astronomy, K\"onigstuhl 17, D-69117 Heidelberg, Germany \\
}

\begin{document}
             
\date{submitted to MNRAS}

\pagerange{\pageref{firstpage}--\pageref{lastpage}}\pubyear{200?}
\maketitle           

\label{firstpage}
             
\begin{abstract}
We use large cosmological N-body simulations  to study 
the subhalo population in galaxy group sized halos. In particular, we look for  
fossil group candidates with typical  masses $\sim$ 10-25\% of 
Virgo cluster but with an  order of magnitude less substructure. 
We examine recent claims that the earliest systems to form are
deficient enough in substructure to 
explain the luminosity function found in fossil groups.
Although our simulations show a correlation between the 
halo formation time and the number of subhalos, 
the maximum suppression of subhalos is a factor of 2-2.5, whereas a factor of 6 is required to match 
fossil groups and galaxies. While the number of subhalos depends
weakly on the formation time,  the slope of the halo substructure velocity function 
does not. The satellite population within Cold Dark Matter (CDM) halos is self-similar at scales 
between galaxies and 
galaxy clusters regardless of mass, whereas current observations show a break in self-similarity
at a mass scale corresponding to group of galaxies.

\end{abstract}


\setcounter{footnote}{1}

\section{Introduction}
\label{sec:intro}

In the current paradigm for cosmological structure formation, 
dark halos collapse from initial gaussian density fluctuations and grow by
accretion and merging in a hierarchical fashion.
A longstanding prediction of the theory is that 
the subhalo (or satellite)  population is self-similar, meaning simply that low mass systems, such as
galaxies are scale-down versions of larger systems, like galaxy clusters.
A galaxy such as the Milky Way is predicted to have nearly the same scaled distribution 
of substructures as a more massive galaxy cluster such as Virgo (Moore et al. 1999, Klypin et al. 1999).
This prediction is tested by using the 
substructure velocity distribution function that expresses the number of sub-halos with circular 
rotational velocity $V_{circ}$ greater than a certain fraction of the circular velocity of 
the parent halo $V_{par}$.

Cosmological numerical simulations confirm the expectations predicting  about 500 satellites with 
$V_{circ}/V_{par}>0.05$ within 500 kpc of the Milky Way. Accounting for the most recent 
discoveries of SLOAN data for fainter galaxies (i.e. Willman et al.
2005, Kleyna et al. 2005) up to date less than 30 satellite galaxies have been observed in the Local
Group including underluminous systems that likely have even lower $V_{circ}/V_{par}$ pointing 
out a  missing satellite problem in the current paradigm for structure formation.
Proposed solutions to this puzzle invoke the reionization of
the universe at high redshift combined with the local injection of energy from star formation
(Bullock et al. 2000, Somerville 2002, Benson et al. 2002).   These mechanisms
may be invoked to suppress galaxy formation in halos below $30 \kms$ reconciling the predictions
with observations  in the Local Group.
Recently, however, D'Onghia \& Lake (2004) have shown that
fossil groups with intermediate mass between the Local group and the 
Virgo cluster are systems with paucity of substructures pointing to a 
missing galaxy problem in higher mass systems.

In fossil groups, a giant E or cD galaxy dominates
with the next brightest group member being at least 2 R-band magnitudes
fainter. The archetype is RXJ1340.6+4018, an isolated, bright 
(M$_{R}=-22.7$) elliptical galaxy at z=0.171, surrounded by a halo of hot
gas and dark matter (Jones et al. 2000).  Its total mass is $\sim 25$\%  
of the Virgo cluster.

With $V_{par}\sim 60$\% of Virgo, the escape energies of their missing 
substructures in the fossil group are 
$\sim10$ times larger than the satellites missing from the Local Group, 
far too large to be explained by proposed suppression mechanisms.

Subsequent observations have shown that fossil groups are clearly virialized objects 
in which a central galaxy is embedded in a X-ray halo with a luminosity 10-60 
Hickson Groups, many of which show only a faint X-ray emission and may be
either evolving systems (Miles et al. 2004) or fake associations; they are more useful than loose groups, 
which are often unvirialized structures 
contaminated by sighting down filaments
(Hernquist et al. 1995).

Recently, however,  Kravtsov et al. (2004) proposed that luminous dwarf spheroidals
in the Local Group descend from downsizing of relatively massive 
($> 10^{9}$M$_{\odot}$) high-z systems by processes such as tidal stripping.    
Again, this is idea may help explaining the lack of substructures in the Local Group (Mayer
et al. 2007) but it is 
unlikely to explain how massive objects ($\sim 10^{10}-10^{11}$ M$_{\odot}$) survive in a
Virgo-size object where the ram stripping is more efficient and  yet are absent in fossil groups
of nearly the same mass.

Is RXJ1340.6+4018 representative of the whole class of fossil groups?
Early work on the fossil groups focused on the magnitude gap in the luminosity function and suggested that it owed 
to early formation times and subsequent merging (Jones et al. 2003, D'Onghia et al. 2005).
However,  little work has been devoted to understand the truly
remarkable feature that nearly every group with a magnitude gap of 2 has a 
substructure function like the Local Group.  The known exceptions are  
 RX J1552.2+2013  
(e.g. Cypriano, Mendes de Oliveira and Sodr\'e 2006) and a cluster of galaxies
discovered recently (Gastaldello et al. 2007)
which are rich enough with substructure that they have been called  "fossil clusters".

The self-similarity of the subhalo population 
resulting from cosmological  N-body simulations is still debated (e.g. De Lucia et al. 2004,
Diemand et al. 2007).  
In this context, Zentner et al. (2005) used semi-analytical models to claim that early formation times lead
to significantly less substructure providing a natural explanation of the lack of galaxies in fossil groups.

These are the issues that we address in this paper. We present 
N-body simulations of galaxy group halos  to clarify the cosmic abundance and
variance of substructure as related to magnitude gaps and formation history.   
The plan of the paper is as follows. In \S 2 we 
present details of the numerical simulations and analysis procedure. \S 3 presents 
the results, whilst 4 summarizes our conclusions.

\section{NUMERICAL METHODS} 
\label{sec:results}

\subsection{Simulations}
We analyze two $\Lambda$CDM simulations with 
cosmological parameters taken from the best-fits for WMAP1 (Spergel et al. 2003) and 
WMAP3 (Spergel et al. 2006). These have  present-day
matter density parameters  $\Omega_m$=0.268 ({\it 0.238} for {\it WMAP3});  cosmological constants
contribution $\Omega_\Lambda$=0.732({\it 0.762}); baryonic contributions 
$\Omega_b,h$=0.044 ({\it 0.042}); and Hubble parameters $h=0.71$({\it 0.73}) ($H_0=100 \ h$ km s$^{-1}$ Mpc$^{-1}$).
The mass fluctuation spectrum has a spectral index $n=1$({\it 0.951}), and is
normalized by the linear rms fluctuation on $8 \rm h^{-1}{\rm Mpc}$, $\sigma_8=$0.9({\it 0.75}). 
The initial   conditions   are   generated  with GRAFIC2  
(Bertschinger 2001).  
All runs started at redshifts sufficiently high to ensure that the absolute maximum density 
contrast is still in the linear regime.

We followed the evolution of $600^3$ particles with masses of $8.98 \times 10^7 \hMsun$ and $8.67
\times 10^7 \hMsun$ for WMAP1 and WMAP3 in a box 90 Mpc (comoving)
on a side using the treecode PKDGRAV (Stadel 2001).
Gravitational interactions between pairs of particles are  softened in comoving coordinates with a  spline softening  length $\epsilon = 1.16$ kpc;  forces are completely  Newtonian at twice this distance.
The particles have individual timesteps $\Delta t_i  = 0.2\sqrt{\epsilon/a_i}$ where 
$a$ is particle's  acceleration.   
The node-opening angle is 
$\theta = 0.7$ after $z = 2$ and $\theta = 0.55$ earlier to provide 
higher force accuracy when the
density is nearly uniform.  Cell moments are expanded to fourth order.

\subsection{Halo identification}

Bound structures with a minimum of 250 particles  at z=0 are identified using the  
Spherical Overdensity (SO)  algorithm (Macci\`o et al. 2007) with a linking length $l$ equal to 0.2 times 
the mean interparticle separation.
For each halo, we fix the center at the most bound particle and find the mass within the
virial radius 
where the spherical overdensity
relative to the critical density for closure is
$\Delta(z=0)\simeq 96.7$ and  $\Delta(z=0)\simeq 93.5$ for WMAP1 and WMAP3 respectively
(Mainini \etal 2003). 
We discard structures with less than 250 particles and
iterate the procedure to convergence. If a  particle is  a potential  member  of two
groups, it  is assigned to the most  massive one.

We select for our analysis all halos with masses in the range 
$M = 1.75-5.25 \times 10^{13} \Msun$, resulting in 28 galaxy group sized halos for WMAP1 and 16 groups for
WMAP3 with $N_{vir}$ between 200,000 and 600,000 particles.

\subsection{Subhalo Identification}

The subhalo population is
identified with SKID (Stadel 2001), which calculates local densities
using an SPH kernel, then moves particles along the density gradient
until they oscillate around a point characterized by  some length $l$.
Then particles are linked together using the friend-of-friend algorithm  with a linking length equal to
$l$. Using a linking length 4 times the gravitational softening SKID algorithm identifies 
the subhalo population inside each halo. Unfortunately by assuming this linking length
the parent halo mass may be underestimated. On the other hand 
using a larger linking length, 10 times the gravitational softening can cure this problem, but then some of
the small Sybila's are missed. Therefore we used a combination of the subhalo
catalogs obtained with these two linking lengths in order to create the
complete catalog of subhalos and to calculate their correct structural parameters.
We select for our analysis only subhalos with $M > 4.5 \times 10^9 \hMsun$.

\subsection{Halo formation redshift}
The assembly history of each halo may be studied using halo catalogs
constructed at various times
during the evolution of the system. For the purposes of our analysis,
we concentrate on the period $0<z<3$. Halos identified at $z>0$
are said to be {\it progenitors} of a $z=0$ system if at least $50\%$
of its particles are found within the virial radius of the
latter. Using this definition we can identify, at all times, the list
of progenitors of a given $z=0$ galaxy group sized halo and track their properties
through time.
For each halo, we use its mass assembly history (MAH)  to determine its formation time.
The MAH is defined using the most massive progenitor 
and is usually
well-fit by:

\begin{equation}
M(z) = M_0 \exp \left [-2\ z \over \left ( 1+ z_f \right )\right ],
\label{eq:MAH}
\end{equation}
with $z$ the redshift, $M_0$ the halo mass at present day  and $z_f$ the formation
redshift, defined as the redshift at which the fitted halo MAH slope assumes the value of 2 (Wechsler et al. 2002).

\begin{figure}
\centering
\psfig{figure=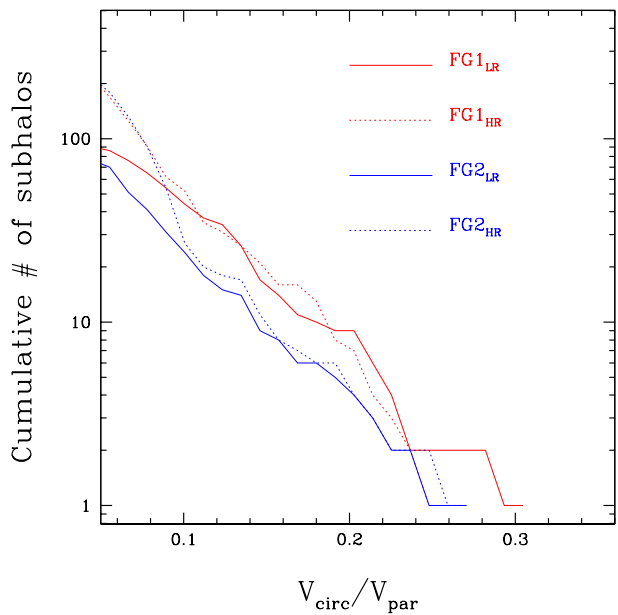,width=250pt}
\caption{Cumulative subhalo velocity distribution function by different 
resolutions for two halos of the sample. 
Solid lines correspond to the lower resolution run and dotted lines to the  
high resolution run of the same halo.}
\label{fig:res}
\end{figure}

\subsection{Numerical Resolution Test}

Only subhalos significantly larger than the gravitational softening
survive the tidal forces of the parent halo.  To test the scale where this  is 
significant, we re-simulated  
two galaxy group sized halos from the WMAP1 simulation
with 8 times higher mass resolution.
The test halos have 3,000,000 particles within the
virial radius at z=0.
Figure \ref{fig:res} shows that the low resolution (solid lines) and the 
high resolution runs (dotted lines) are in fairly good agreement 
for subhalos above our adopted resolution limit of  $V_{circ}/V_{par} =  0.1$. 

\subsection{Fossil group sample selection criterion}

We aim to select fossil groups in our sample of galaxy group sized halos.
Ideally we want to determine a distribution of magnitude gaps and its relationship to the 
subhalo population in our sample. Since luminosity and magnitudes are not defined in our simulations,
the identification of a magnitude gap between the first and second brightest member
must be translated to a circular velocity gap.
If the Faber-Jackson (1976) relationship holds, then a 2 magnitude gap
in the luminosity function corresponds
to a circular velocity ratio $V_{circ}^{2nd}/V_{circ}^{1st} \sim 0.6$, while a gap of 2.5 magnitudes would be  
$V_{circ}^{2nd}/V_{circ}^{1st} \sim 0.55$.
In observed systems, the brightest halo lies at the center of the X-ray emission as well
as at the centroid of the redshift distribution, but its velocity
dispersion is significantly less than the overall dispersion of the group, or equivalently, its X-ray
temperature.  The first brightest galaxy in 
our simulations is subsumed into the parent halo, so we need to calculate the 
ratio of $V_{circ}^{2nd}/V_{par} $ that characterizes a fossil group.
In the case of RXJ1340.6+4018, its luminosity function has a magnitude gap of 2.5, corresponding to the ratio of 
$V_{circ}^{2nd}/V_{circ}^{1st}= 0.55$ and $V_{circ}^{1st}/V_{par} \sim 0.65$, so 
that $V_{circ}^{2nd}/V_{par} \sim 0.35$. 
We adopt  $V_{circ}^{2nd}/V_{par} < 0.35$ as a conservative criterion for a group with an appropriate 
magnitude gap.  We note that 
the Local Group and Centaurus A group also meet this criterion and have substructure functions comparable
to RXJ1340.6+4018.
Subhalo mass determination in collisionless N-body simulations is uncertain. 
When  dissipation is included in simulations, the subhalos number increases 
because the substructures become more robust and can survive 
to the very center of the parent halos  (Nagai \& Kravtsov 2005, Macci\`o et al. 2006a).  Hence we measured 
the circular velocities of substructure at the time of {\it infall}  into
the larger halos (see also Strigari et al. 2007).  After the subhalo falls into the parent halo it loses
mass and its circular velocity decreases over time. Henceforth we 
compare the subhalo circular velocity at infall time with the observational data, not the
subhalo circular velocities measured at present time.

\begin{figure}
\centering
\psfig{figure=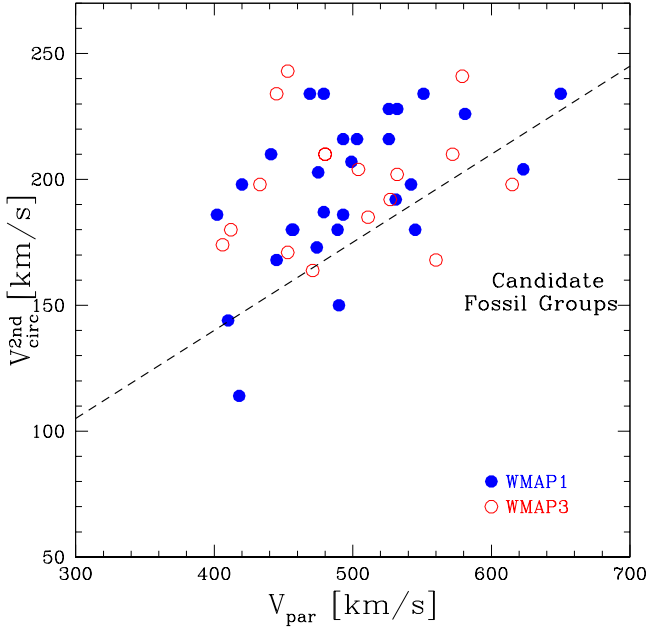,width=250pt}
\caption{The circular 
velocity of the second most massive subhalo versus the circular velocity of the parent halo
in the sample of group sized halos for the two set of simulations with WMAP1 (filled symbols)  and WMAP3
(empty symbols) cosmological parameters.  
The dashed line corresponds to the dividing line for a velocity gap comparable to that observed in 
fossil groups.}
\label{fig:cos}
\end{figure}

Figure \ref{fig:cos} shows the circular 
velocity of the second most massive subhalo compared to that of  the parent halo
for our simulated groups. The groups with $V_{circ}^{2nd}/V_{par} <0.35$ are selected as 
candidate fossil groups. Only ~18\% of groups qualify, in fair agreement with the observed rate 
(Jones et al. 2003, van den Bosch et al. 2007).  
But, does their substructure function look like that one of the fossil group RXJ1340.6+4018?

\section{Results}
\label{sec:res}
\subsection{Substructure function for different cosmological parameters}

\begin{figure}
\centering
\psfig{figure=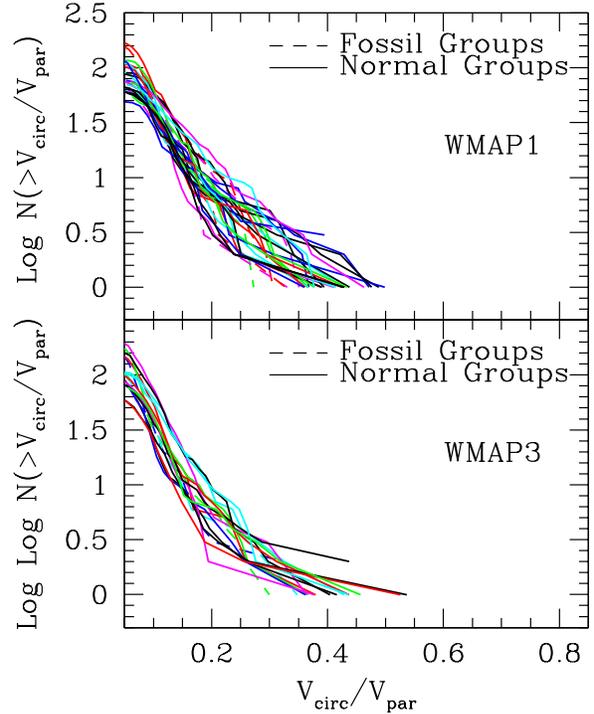,width=250pt}
\caption{Cumulative substructure function  for the
 group sized halos for the different cosmological parameters: WMAP1 in the top panel  
 and WMAP3 in the bottom.  Solid lines  
correspond to normal group sized halos, while dashed lines are the 
candidate fossil groups.}
\label{fig:VDF}
\end{figure}

We define a system to have a {\it galaxy-like} substructure function when it rescales
to that of the Milky Way
or equivalently RXJ1340.6+4018 (see e.g. Figure 1 in D'Onghia \& Lake 2004). 
The opposite behavior would be a system with a {\it cluster-like} 
behavior,  e.g. a rescaled  Virgo cluster.
The cumulative subhalo velocity distribution function of  the halo  
sample is shown in  Figure \ref{fig:VDF}. Note that in this figure we remove the most massive object (i.e. the parent which should include the most massive galaxy) 
from the cumulative substructure velocity distribution.

The top (bottom) panel 
is the simulations with
WMAP1 (WMAP3) cosmological parameters. 
The candidate fossil groups  are shown by 
dashed lines  against a background of solid lines for the normal groups.
The panels show that the principal difference between the two 
realizations is that there are fewer group-sized halos with WMAP3
parameters.  There is no  statistically significant difference between the total number of subhalos
with $V_{circ}/V_{par} > 0.1$ for candidate fossil groups and normal groups in 
the two realizations.  The cumulative substructure   velocity distribution functions also  
show similar trends.

\subsection{Comparison to observational data}

\begin{figure}
\centering
\psfig{figure=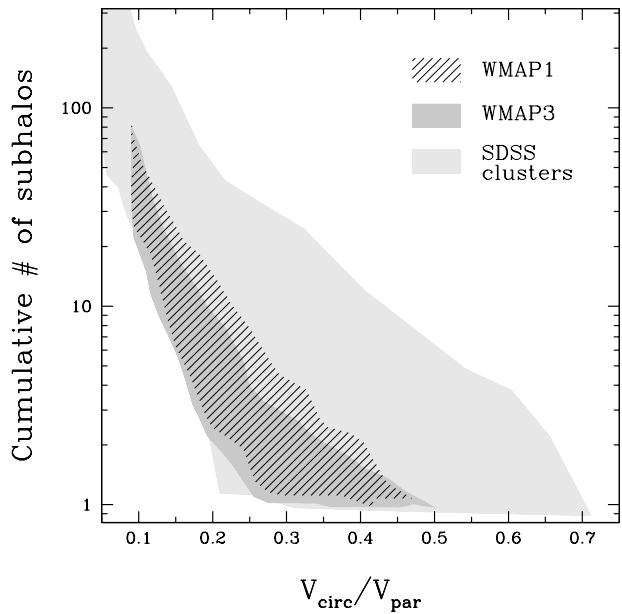,width=250pt}
\caption{ Observed cumulative substructure function within
  galaxy clusters inferred from SDSS data, light gray area (Desai et al. 2004) compared to CDM
  predictions for all the galaxy group sized halos. The areas  represent N-body simulation results
  for galaxy groups extracted from WMAP1 (diagonal texture) and WMAP3 simulations (gray
  area).The substructure function is the number of objects with
  velocities greater than a fraction of the parent halo's velocity.}
\label{fig:desai}
\end{figure}

\begin{figure}
\centering
\psfig{figure=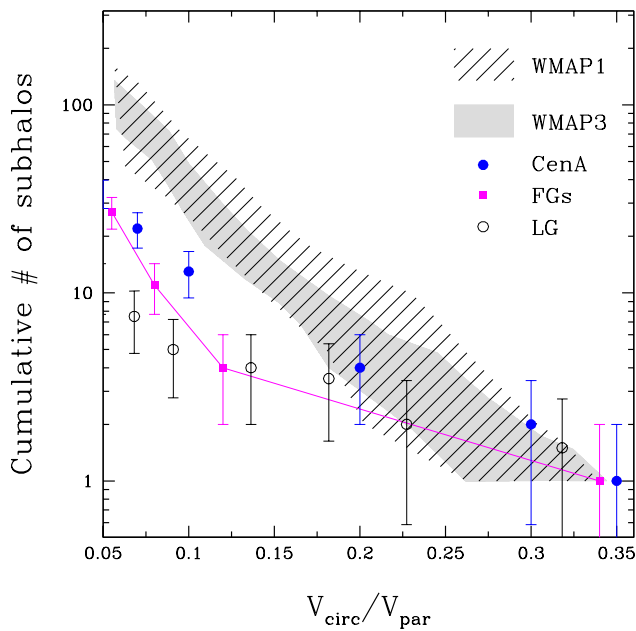,width=250pt}
\caption{ Observed cumulative substructure function within
  RX J1340.6+4018, Virgo cluster, the Local Group and Centaurus A compared to CDM
  predictions. The shaded area and the  represent N-body simulation results
  for candidate fossil groups extracted from WMAP1 (diagonal texture) and WMAP3 simulations (gray
  area).}
\label{fig:obs}
\end{figure}

The range of cumulative substructure   velocity distribution functions for all the 
galaxy group sized halos  is shown  
in Figure \ref{fig:desai} as compared to the observed ones for galaxy clusters 
inferred from SDSS data (Desai et al. 2004). The agreement between predictions and observations
is encouraging, 
although the observed cumulative distribution
exhibits a broader range for $V_{circ}^{2nd}/V_{par}$
from 0.25 to 0.8, whereas the second most massive subhalo in the simulations is never larger than
50\% of the parent halo. The larger spread might owe 
to uncertainties in the estimate of the parent's halo velocity dispersion from a relatively 
small number of member velocities.   
The trend of the cumulative subhalo velocity function in simulations
agrees well with the one inferred from galaxy clusters. Note that 
 Figure \ref{fig:desai} shows that there are galaxy clusters with the second most massive member 
 having $V_{circ}\sim$ 25-30\% of the parent 
halo, showing evidence for magnitude gaps in SDSS data.

Figure \ref{fig:obs} shows 
the range of cumulative subhalo velocity distribution functions for candidate 
fossil groups in simulations, marked by shaded areas,  as compared to the observed ones for 
RXJ1340.6+4018, the Local Group and 
Centaurus A (D'Onghia \& Lake 2004 for a compilation and reference therein). 
The Local Group uses data 
from Mateo et al. (1998), Odenkirchen et al. (2001) and Kleyna et al. (2005); we
used the results of Kazantzidis et al. (2004) to convert
three-dimensional velocity dispersion into circular velocity.
We considered the number of satellites per central
galaxy instead of considering the Local Group as a whole. This makes the
comparison with simulations straightforward (see Macci\`o et al. 2006b
for more details). 

Note that while we find  systems with appropriate gaps in velocity between the second
most massive subhalo and the most massive one, the cosmic
variance in CDM simulations is insufficient to explain the cumulative number of subhalos
with circular velocity less than 20\% of observed systems (e.g. $V_{circ}/V_{par} < 0.2$).

Numerical simulations performed with  
WMAP1 and WMAP3 parameters  give almost identical results; the narrower
region covered by the velocity distribution function in WMAP3  owes to the lower number
of candidate fossil groups. These systems populate the lower region of the 
shaded area shown in Figure \ref{fig:desai}. The slope of the function 
agrees well with the one inferred from normal groups and galaxy clusters.
Yet, both WMAP1 and WMAP3 models over-predict the total
number of substructures relative to Centaurus A and RXJ1340.6+401. Especially for 
this latter one the discrepancy in the velocity range of subhalos with circular velocity larger
than 10\% is a factor of six.
Can this discrepancy owe to uncertainties in the estimate
of the total mass (circular velocity) of the parent halo of RXJ1340.6+401?
If the total mass is lower than estimated \footnote{The velocity dispersion for the sistem is quoted $\sigma \sim 380$ 
km/s with great uncertainties due to the low number of galaxy members (Jones et al. 2000).}
the points in Figure  \ref{fig:obs}
may shift on the right leading to a fair agreement with the expectations (dashed area).
However the X-ray temperature 
is $\sim$ 1 KeV for 
the archetype fossil group, meaning a circular velocity  of the parent halo
at least 580 km/s, so the points can't be shifted to the right to obtain
a better agreement.

Also, dynamical friction and merging are not a general solution to the discrepancy. 
These dynamical effects are included in the full numerical simulations. 
Any specific substructure function evolves in the same way by dynamical friction and merging 
independent of the parent mass (D'Onghia \& Lake 2004).
Dynamical friction can be important in promoting the merger of the largest objects in less than one Hubble time 
but dynamical friction alone will not create substructure functions 
that are different for different parent masses. Clearly galaxies could have had a long time to evolve 
by dynamical friction, 
but this is also included in the simulations and does not affect the substructure 
function below $V_{circ}/V_{par}$ of 0.2.

\begin{figure}
\centering
\psfig{figure=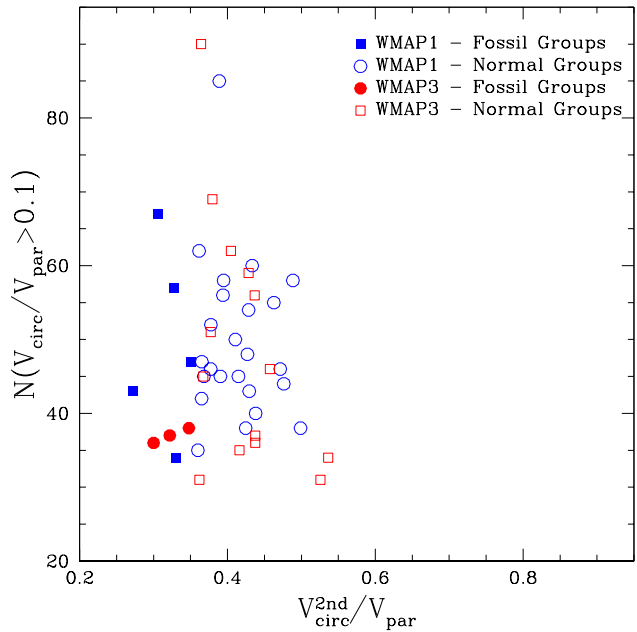,width=250pt}
\caption{Total number of subhalos with circular velocity
larger than 10\% of the parent circular velocity is plotted against the circular velocity  
of the second most massive subhalo normalized to the parent halo velocity}
\label{fig:Nv2vp}
\end{figure}

Figure \ref{fig:Nv2vp}
shows that the total number of substructures with ($V_{circ}/V_{par}>0.1$) 
in our simulations  is independent of any velocity gap: $V_{circ}^{2nd}/V_{par}$.
Yet, we don't find a halo population as poor  in substructure as  RXJ1340.6+4018 and the Local Group.

\subsection{Substructure abundance and formation redshift}

Recently, using semi-analytic models, Zentner et al. (2005) have 
highlighted the possibility that early forming halos may 
have significant less substructures, providing a potential 
explanation of the lack of substructures in the observed fossil groups.

\begin{figure}
\centering
\psfig{figure=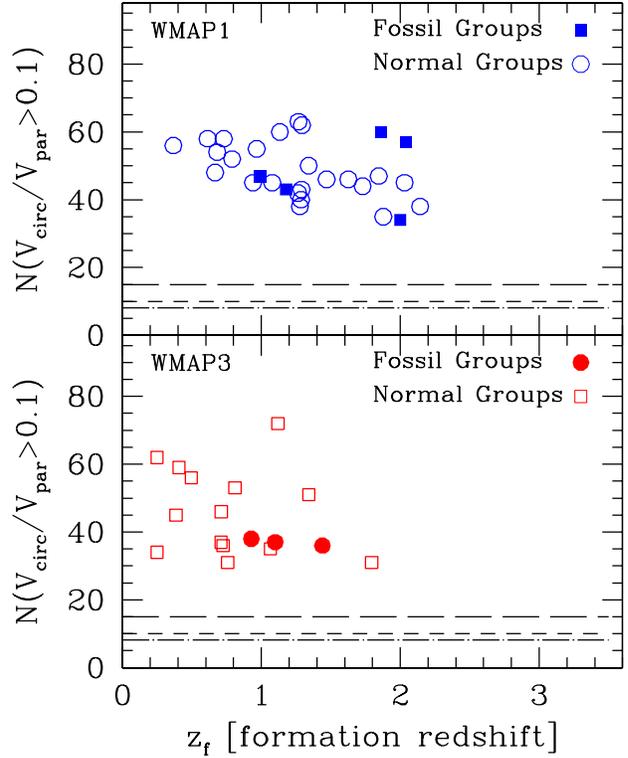,width=250pt}
\caption{Total number of satellites with circular velocity
larger than 10\% of the parent circular velocity
 versus the formation redshift of the parent halo. Solid (red) circles mark 
normal group sized halos, open (blue) squares the candidate fossil group halos. 
Top and bottom panel show results for the WMAP1 and WMAP3 simulations, 
respectively. The three lines on the bottom of the plot show 
observational results inferred for Centaurus A, RXJ1340.6+4018 and the 
Local Group.}
\label{fig:N01}
\end{figure}

We explore this claim in Figure \ref{fig:N01}, where the total number of 
subhalos with
circular velocity larger than 10\% of the parent halo is shown  as a function of
the formation redshift ($z_f$) of the parent halo and the two sets of cosmological parameters. Both panels
 show that there is an anticorrelation
between the halo formation redshift and the total number of substructures. Halos that form
earlier exhibit a lower number of satellites 
within a factor two for a cosmology with WMAP1 parameters  and slightly larger and with more spread
adopting WMAP3 parameters. 
We tested also the distribution of the halo formation redshift 
of all the halos selected in the WMAP1 and WMAP3 simulations. 
Fossil groups and normal groups in the range of mass of $1-5 \times 10^{13}$M$_{\odot}$ here adopted 
tend to form in a narrow range of redshift centered
around z=1.5 and z=0.9 for cosmologies with WMAP1 and WMAP3 parameters, respectively. The 
candidate fossil group halos tend to form slightly earlier than the normal groups ones, even if the
sample of fossil groups is limited.

Is this the solution to the missing galaxy question in RXJ1340.6+4018?
In Figure \ref{fig:N01} we draw three additional horizontal lines 
tracing the total number of observed galaxies with circular velocity larger than 10\% of the parent 
halo velocity ($V_{circ}/V_{par} \ge 0.1$)
for the galaxy group Centaurus A, RXJ1340.6+4018 and the Local Group.
While the early formation redshift 
can reduce on average of a factor two the total number of satellites in simulations, 
the observations show a discrepancy greater than a factor of five. 
{\it We conclude that the early formation time per se can reduce but not 
reconcile the discrepancy between the observed total number of satellites to 
the number predicted by numerical simulations in galaxy groups.}

Is the anticorrelation between total number of satellites and 
redshift formation exhibited in Figure 6 in conflict with the 
previous finding of self-similarity of the subhalo population of dark
matter halos?  In order to assess this effect we analyzed separately for each halo  
the {\it normalization} of the cumulative velocity distribution function
and its slope as a function of the formation redshift. 
The substructure function  of a all halos is known to be approximated by  the
following 
power law:

\begin{equation}
N \left ( > \frac{V_{circ}}{V_{par}} \right )=V_0 * \left (V_{circ} \over V_{par} \right )^{\gamma}
\end{equation}

with $\gamma=-2.75$ over a wide 
range of mass  and cosmological models (Klypin \etal 1999, Klypin \etal 2003).
For each halo we fit the power law in the range of  
$ V_{circ}/V_{par}$ between 0.1 and 0.3 in order to 
to avoid to bias the fossil group  candidates  in respect to the normal 
group halos, since the first ones have, by definition 
a lack of subhalos for $V_{circ}/V_{par}>0.35$. 
During the fitting procedure the normalization $V_0$ and the slope $\gamma$ 
are treated as free parameters.
The values and associated uncertainties of  $V_0$ and $\gamma$ are determined via a $\chi^2$ minimization
procedure using the Lavenberg \& Marquart method. 

\begin{figure}
\centering
\psfig{figure=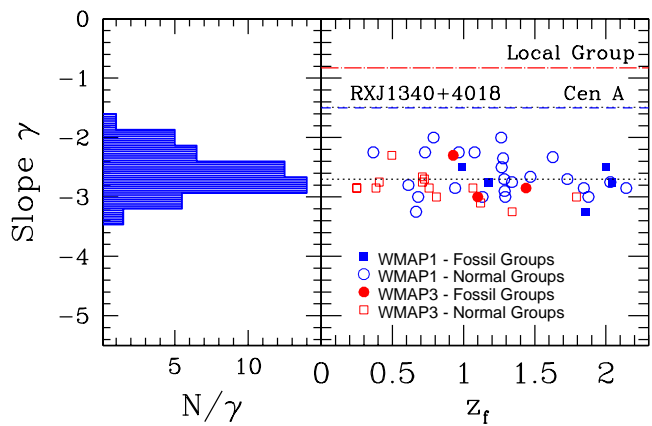,width=250pt}
\caption{{\it Left Panel}. The slope distribution of all the group sized halos of the samples.  
{\it Right Panel}. Slopes of the subhalo function versus 
  formation redshift for all the halos in the two simulations (WMAP1/WMAP3). 
  The slopes inferred for the observed SDSS galaxy cluster sample,  
 the Local Group, Centaurus A and RXJ1340.6+4018 are also indicated for comparison.}
\label{fig:slopes}
\end{figure}

The slope distribution of all the halos selected at z=0 is shown by the 
histogram in the left panel of Figure \ref{fig:slopes}. 
The panel on the right of the Figure \ref{fig:slopes} shows the 
lack of correlation between the slope $\gamma$
of the substructure function  of each halo and its formation redshift $z_f$.

{\it We conclude that only the  normalization of the cumulative velocity 
distribution function of satellites anticorrelates with the formation time, 
the slope does not}. This result highlights that non-linear structures in CDM models 
are self-similar, independent of the formation time and insensitive to the difference
in cosmological parameters used in the two simulations. 
Thus, while the
slope of the cumulative substructure function is set by the cosmological model (setting the
self similarity of dark halos), its normalization depends on
the local environment and on the formation history of the halo.

We draw additional horizontal lines in Figure \ref{fig:slopes} showing the slope inferred for
the Local Group, RXJ1340.6+4018  and Centaurus A and SDSS galaxy clusters 
(the middle two are so close that they
are hard to resolve). The SDSS galaxy cluster slope quoted by Desai et al. (2004) is $\gamma=-2.614 \pm 0.042$
in fair agreement with numerical simulations. This shallower value quoted for SDSS data
is because it is estimated over the entire range of $V_{circ}/V_{par}$, whereas
the slope of the simulations is obtained in a narrower range (for 0.1$< V_{circ}/V_{par} <$0.3). 
However, the observed slopes 
of galaxy groups like Centaurus A and RXJ1340.6+4018 are substantially lower.

\section{Summary}
\label{sec:conclusion}

RXJ1340.6+4018 has a typical mass of 25\% of the Virgo cluster mass, but it 
exibits a substructure population similar to the rescaled version  of the  
Local Group. 
We used numerical simulations of group sized halos to examine  
the formation of systems with paucity of substructures like 
the fossil system RXJ1340.6+4018.

Our results show that:

\begin{itemize}
\item 
In the mass range 1-5$\times 10^{13}$M$_{\odot}$ almost 18\% of the group 
sized halos are candidate fossil groups according to the definition of 
velocity gap, but not a single system 
has a shallow cumulative substructure function similar to that one exhibited 
from the Local Group or RXJ1340.6+4018. 

\item 
Current observations of galaxy groups with a velocity gap
 exhibit a paucity of substructures with a substructure function like that of the Local Group,
whereas the total number of substructures predicted from CDM models is almost 
constant and independent of any gap in velocity. 

\item The total number of substructures identified at the present time does indeed correlate
with the formation time of the halo, as found in semi-analytics work. 
However, the early formation time reduces the average number of substructures by 
a factor two, while  a factor of six is needed to reconcile the
predictions with the observations. We showed that only the  normalization of the cumulative velocity function  of 
satellites depends on the formation time, whereas the slope does not.  This seems intuitive as the fluctuations in
CDM models are gaussian with random phases.  While the early formation of a parent halo from a large
scale wave will bias
the formation of substructures, it should not and indeed does not tilt the smaller scale spectrum.

\item 
We find that  the cumulative velocity distribution function of substructures is self-similar 
in CDM models, with  the  slope being the same in the WMAP1 and WMAP3 realizations 
and they agree with previous results on halos with galaxy masses
and galaxy clusters. Yet, only the slope of the cumulative velocity function of substructures in 
the SDSS galaxy cluster sample agrees with the predictions of the numerical simulations, while 
the slopes of groups of galaxies like the Local Group, Cenaturus A or 
the RXJ1340.6+4018 are  significantly shallower.

\end{itemize} 
We explored the effects of cosmic variance on the cumulative substructure function and find that the deviation
of the model predictions from the observations begins for subhalos with circular velocity 
larger than a 20\% of the parent halo circular velocity.
For a typical galaxy group sized halo with circular velocity  of $450 \kms$, a missing subhalo
in the observations has a circular velocity of  $\sim 80 \kms$.   

Physical mechanisms invoking 
reionization or  feedback processes  from satellites 
are already stretching
to suppress halos at $30 \kms$. 
Here we showed that invoking a  earlier  formation time for these objects may alleviate
the problem but cannot reproduce the observed number of visible galaxies
in groups of galaxies with intermidiate mass between galaxies and poor cluster of galaxies.

If future spectroscopic luminosity function for X-ray emitting groups 
in the range of mass intermidiate between the Local group and the Virgo cluster  produces
a larger sample of groups with paucity of substructures  this class of objects will remain 
an interesting  challenge for the current model
of structure formation.

\section*{Acknowledgments} 

The numerical simulations were performed on zBox2 supercomputer at the 
University of Zurich; (http://www-theorie.physik.unizh.ch/$\sim$dpotter/zbox2/). 
E.D. thanks highlighting discussions with Daniele Pierini, Maurilio Pannella and Stefano Zibetti 
on observational data. E.D. is financed by a EU Marie Curie Intra-European Fellowship under contract
MEIF-041569.


\label{lastpage}
\end{document}